\documentclass[smus]{snow2e}
\usepackage{graphics}
\begin{document}
\def\pT{p_T^{\phantom{7}}}
\def\MW{M_W^{\phantom{7}}}
\def\ET{E_T^{\phantom{7}}}
\def\bh{\bar h}
\def\lm{\,{\rm lm}}
\def\lo{\lambda_1}
\def\lt{\lambda_2}
\def\pslt{p\llap/_T}
\def\to{\rightarrow}
\def\Re{{\cal R \mskip-4mu \lower.1ex \hbox{\it e}}\,}
\def\Im{{\cal I \mskip-5mu \lower.1ex \hbox{\it m}}\,}
\def\SU{SU(2)$\times$U(1)$_Y$}
\def\te{\tilde e}
\def\tu{\tilde u}
\def\tmu{\tilde \mu}
\def \tt{\tilde t}
\def\tl{\tilde l}
\def\tb{\tilde b}
\def\tevst{TeV$^*\ $}
\def\tf{\tilde f}
\def\mhf{m_{1/2}}
\def\ttau{\tilde \tau}
\def\tg{\tilde g}
\def\tga{\tilde \gamma}
\def\tnu{\tilde\nu}
\def\tell{\tilde\ell}
\def\tq{\tilde q}
\def\tst{\tilde t}
\def\tw{{\tilde \chi}^\pm}
\def\tz{{\tilde \chi}^0}
\def\cmsec{{\rm cm^{-2}s^{-1}}}
\def\sgn{\mathop{\rm sgn}}
\hyphenation{mssm}
\def\ds{\displaystyle}
\def\ts{${\strut\atop\strut}$}

\catcode`\@=11 
\newcommand{\lsim}{\mathrel{\mathpalette\@versim<}}
\newcommand{\gsim}{\mathrel{\mathpalette\@versim>}}
\def\@versim#1#2{\vcenter{\offinterlineskip
 \ialign{$\m@th#1\hfil##\hfil$\crcr#2\crcr\sim\crcr } }}
 \def\cropen#1{\crcr\noalign{\vskip #1}}
\def\crr{\cropen{1\jot }}
\def\eslt{\mathrel{\m@th\mathpalette\c@ncel E_T}}
\catcode`\@=12 

\title{Summary of the Supersymmetry Working Group}

\author{Jonathan  Bagger\\
\textit{Department of Physics and Astronomy,
Johns Hopkins University,
Baltimore, MD 21218}\\
Uriel  Nauenberg\\
\textit{Department of Physics,
University of Colorado,
Boulder, CO 80309}\\
Xerxes  Tata\\
\textit{Department of Physics and Astronomy,
University of Hawaii,
Honolulu, HI 96822}\\
Andrew  White\\
\textit{Department of Physics,
University of Texas at Arlington,
Arlington, TX 76019}\\
\\ GROUP MEMBERS \\[2mm]
\begin{tabular}{lll}
J. Amundson (Wisconsin) & G. Anderson (FNAL) & H. Baer (FSU) \cr
J. Bagger (Johns Hopkins) & A. Barbaro-Galtieri (LBNL) & R.M. Barnett (LBNL)
\cr
A. Bartl (Vienna) & D. Burke (SLAC) & R. Cahn (LBNL) \cr
W.B. Campbell (Nebraska) & C.-H. Chen (UC Davis) & G. Cleaver (OSU) \cr
F. Cuypers (PSI) & M.N. Danielson (Colorado) & K. De (Texas, Arlington) \cr
M. Diwan (BNL) & B. Dobrescu (Boston) & M. Drees (Wisconsin) \cr
R. Dubois (SLAC) & S. Fahey (Colorado) & J.L. Feng (LBNL) \cr
K. Fujii (KEK) & E. Goodman (Colorado) & J.F. Gunion (UC Davis)\cr
I. Hinchliffe (LBNL) & T. Kamon (Texas A\&M) & G.L. Kane (Michigan) \cr
B. Kayser (NSF) & C. Kolda (IAS) & S. Lammel (FNAL) \cr
J. Lykken (FNAL) & S. Manly (Yale) & S.P. Martin (Michigan) \cr
T. Moroi (LBNL) & S. Mrenna (Argonne) & R. Munroe (FSU) \cr
U. Nauenberg (Colorado) & M. Nojiri (KEK) & D. Norman (Texas A\&M) \cr
F. Paige (BNL) & M.E. Peskin (SLAC) & D. Pierce (SLAC) \cr
J. Pilcher (Chicago) & L. Poggioli (CERN) & A. Seiden (UC Santa Cruz) \cr
M. Shapiro (LBNL) & A. Skuja (Maryland) & M. Sosebee (Texas, Arlington) \cr
J. S\"oderqvist (KTH) & X. Tata (Hawaii) & S. Thomas (SLAC) \cr
D.L. Wagner (Colorado) & J.D. Wells (SLAC) & J.T. White (Texas A\&M) \cr
A. White (Texas, Arlington) & B. Williams (Colorado) & B. Wright (North
Carolina) \cr
Y. Yamada (Wisconsin) & W.-M. Yao (LBNL) &  \cr
\end{tabular}}
\vspace{2.0in}
\maketitle

\thispagestyle{empty}\pagestyle{empty}
\lefthyphenmin=2
\righthyphenmin=3

\begin{abstract}
We summarize the results obtained by the Supersymmetry Working
Group at the 1996 Snowmass Workshop.
\end{abstract}

\section{INTRODUCTION}

Supersymmetry (SUSY) is a novel symmetry which relates the
properties of bosons and fermions.  It implies that the
elementary particles come in pairs, with the same masses
and internal quantum numbers, but with spins differing by
one-half unit of angular momentum.  Such particles do
not exist, so SUSY must be broken (a bosonic electron
of mass 0.511 MeV could not have escaped detection).

Interest in SUSY phenomenology stems from the fact that
in a supersymmetric theory, the masses of elementary scalar
particles remain stable under radiative corrections -- even
if supersymmetry is softly broken.  For the case of the
standard model (SM), this suggests that the scale of SUSY
breaking should be comparable to the weak scale, $\sim$
250 GeV.  In this case, the unobserved SUSY partners must
have masses smaller than $\sim$ 1 TeV, and they should be
copiously produced at the next generation of colliders, if
not before.

The Minimal Supersymmetric Standard Model (MSSM) is the direct
supersymmetrization of the SM.  It is an $SU(3) \times SU(2)
\times U(1)$ gauge theory, with three generations of quarks and
leptons, together with their spin-zero partners, the squarks and
sleptons.  The electroweak symmetry-breaking sector contains two
$SU(2)$-doublet scalar fields, along with their spin-$\frac{1}{2}$
partners, the Higgsinos.  The remaining fields of the MSSM are
the spin-$\frac{1}{2}$ Majorana gauginos, which transform as the
adjoint representation of the SM gauge group.

Aside from (model-dependent) mixing effects, the gauge interactions
of all the sparticles are completely fixed by gauge invariance.
These interactions largely determine the phenomenology of the
MSSM, and allow for rather robust predictions.  Just as in the SM,
the masses of the fermions arise via Yukawa interactions, which
are now contained in a superpotential.  The superpotential function
also contains a supersymmetric Higgs mass term, which is conventionally
denoted by $\mu$.

In contrast to the SM, the most general gauge-invariant
Lagrangian for the MSSM contains renormalizable interactions which
violate baryon and/or lepton number.  These interactions lead to
squark-mediated weak-scale proton decay, which is clearly
unacceptable.  Typically, these interactions are {\it assumed}
to be forbidden by a $Z_2$ symmetry called $R$-parity, which
is defined to be $+1$ for ordinary particles, and $-1$ for
their superpartners.

$R$-parity conservation implies that supersymmetric particles must
be pair-produced in collisions of ordinary particles, and that they
must decay into other supersymmetric particles.  It also implies
that the lightest SUSY particle (LSP) is stable.  We note that
it is possible to construct phenomenologically viable models
where $R$-parity is not conserved.  The phenomenology of such
models is significantly different from that of the MSSM, and is
discussed elsewhere in this report.

The absence of Bose-Fermi pairs of the same mass implies that
supersymmetry must be broken.  It is fair to say that there is,
as yet, no compelling mechanism for supersymmetry breaking.  For
phenomenological purposes, however, it is sufficient to parametrize
the effects of supersymmetry breaking by introducing a set of soft
SUSY-breaking operators which are consistent with the SM
symmetries.  These operators do not reintroduce the divergences
which destabilize the scalar masses.  They also play
an important role in achieving the spontaneous breakdown of
electroweak gauge invariance.

Once electroweak symmetry is broken, gauginos and Higgsinos of the
same charge mix to form the mass eigenstates known as the charginos
and neutralinos.  The gluinos, being the only color-octet fermions,
do not mix with any other particles, and so are mass eigenstates.
Left-right sfermion mixing, which is proportional to the corresponding
{\it fermion} mass, is negligible except for the third generation.
Therefore $\tf_L$ and $\tf_R$, the superpartners of the chiral fermions
$f_L$ and $f_R$ of the first two generations, are essentially mass
eigenstates, whereas there is always substantial mixing between the
left- and right-handed top squarks.  (The mixing between left-
and right-handed bottom squarks and tau sleptons depends on model
parameters.)

The MSSM thus contains two Dirac charginos, four Majorana neutralinos,
a color-octet gluino, two spin-zero squarks and charged sleptons of each
flavor, and a sneutrino for each lepton family.  The Higgs sector
contains three neutral Higgs states plus a pair of charged Higgs
particles.  As noted above, sparticles decay into other sparticles
until the cascade terminates with the stable LSP.  The favored LSP
candidate is the lightest neutralino, which is weakly interacting
and escapes experimental detection.  Thus missing energy and missing
momentum are the canonical signatures for SUSY, provided $R$-parity
is conserved.

The soft SUSY-breaking operators have been classified in
Ref.~\cite{GIRARDELLO}.  They consist of scalar and gaugino
masses, together with trilinear and bilinear scalar interactions
(the so-called $A$- and $B$-terms).  If one neglects flavor
mixing in the supersymmetric sector, one finds an independent
mass term for each $SU(3) \times SU(2) \times U(1)$ matter,
Higgs and gaugino multiplet (for a total of $15+2+3$ masses).
One also finds a soft trilinear scalar coupling for every
superpotential interaction.  With the most general flavor
structure, the sfermion mass parameters are replaced by sfermion
mass matrices, and the number of trilinear couplings is greatly
increased.  The generic model with the minimal particle content
contains over one hundred soft SUSY-breaking parameters (and
even more if $R$-parity is not conserved).  This makes general
SUSY phenomenology intractable.

In the future, a theory of these soft parameters will emerge
from a deeper understanding of the mechanism which underlies
supersymmetry breaking.  For the present, however, one is forced
to take a more practical approach and find some other way to reduce
their number.  A reasonable strategy is to postulate symmetries
which hold at some high energy scale.  For example, one might suppose
that there is an $SU(5)$ grand unified symmetry, where each family
falls into one of two $SU(5)$ multiplets.  Then, in one common
scenario, there are just two sfermion masses for each family,
and only one gaugino mass.

As usual in a unified theory, these $SU(5)$ mass relations hold at
the unification scale.  The physical masses are found by evolving
the mass parameters to the weak scale.  The evolution equations
depend on the interactions of the individual sparticles, so a rich
pattern of physical sparticle masses can emerge.  For example,
in the above scenario one finds that the weak-scale gaugino mass
parameters are related by the well-known gaugino mass unification
relation,
\begin{displaymath}
\frac{3M_1}{5\alpha_1}=\frac{M_2}{\alpha_2}=\frac{m_{\tg}}{\alpha_s}.
\end{displaymath}
Of course, the precise physical mass spectrum depends on the
assumptions we make about the physics at the high scale.  We
will return to this point in the next section, where we discuss
the canonical model that we use in most of this study, as well
as possible variations.

The goals of our study are embodied in the following questions:

\begin{itemize}

\item Can one identify a signal for new physics in a supersymmetric
context?

\item Once a signal is identified, can one tell that the new physics
is supersymmetry?

\item Having identified the new physics as supersymmetry, can one
distinguish between various models and actually measure the underlying
parameters?

\end{itemize}

The future experimental facilities which were included in our charge,
and for which we studied these issues, were as follows:

\begin{itemize}

\item TeV33, a high luminosity upgrade of the Tevatron collider,
operating at $\sqrt{s}=2$~TeV, with an integrated luminosity
$\sim$ 30 fb$^{-1}$,

\item LHC, where studies were primarily done for the ``low luminosity''
option of 10 fb$^{-1}$/year, and

\item NLC, an $e^+e^-$ collider operating $\sqrt{s}= 0.25- 1.5$~TeV,
with a design luminosity of up to 50 fb$^{-1}$/year, depending on
the energy.
\end{itemize}

Our group divided itself into four subgroups with the following
conveners:

\begin{itemize}
\item Theory (H. Baer and J. Lykken)
\item TeV33 (K. De and S. Lammel)
\item LHC (A. Bartl and J. S\"oderqvist)
\item NLC (K. Fujii and D. Wagner).
\end{itemize}

\noindent
The summaries of each of the subgroups appear elsewhere in these
proceedings.

In Sec.~II we outline the assumptions which underlie what has
come to be known as the minimal SUGRA framework, and which is
used as a canonical model for the experimental studies.  We also
explore possible variations and discuss their implications, as
elucidated by the Theory Subgroup.  In Sec.~III we briefly
review the SUSY reach of the various facilities, as documented
in the literature prior to this Snowmass Workshop.  In Sec.~IV
we outline strategies that can be used in future experiments to achieve
the goals of this study, and we discuss highlights of the new
results that were obtained at the Workshop.  We conclude in
Sec.~V with a comparison of the capabilities of the three
experimental options, assuming that weak-scale supersymmetry is
indeed realized in nature.

\section{MODEL CONSIDERATIONS}

As we have just seen, the minimal supersymmetric extension of
the SM contains over one hundred free parameters.  Clearly,
to proceed further, one must make assumptions.

Most recent phenomenological analyses have been done within
the so-called minimal SUGRA framework, which we also adopt
for most of our analysis.  Here, we briefly review the
assumptions that underlie this scenario, and then we go on
to discuss other theoretical models that were considered at
this Workshop, primarily by the Theory Subgroup.

\subsection{The Minimal SUGRA Model}

Various difficulties in constructing phenomenologically
viable models with spontaneously broken weak-scale SUSY led
to the development of the so-called geometric hierarchy
models, where SUSY is broken in a hidden sector at a scale
$M_{\rm SUSY}$.  In these models, the interactions between the
hidden and observable sectors are suppressed by some large
scale, $M$.  The effective scale of SUSY breaking in the
observable sector is $M^2_{SUSY}/M$, which can be comparable
to the weak scale even if $M_{\rm SUSY}$ is much larger.

Supergravity provides a particularly attractive realization
of this idea, in which case $M \sim M_{\rm Planck}$.  In the
supergravity framework, SUSY is promoted to a local symmetry,
and the resulting model necessarily includes gravity.  The gravitino
typically acquires a weak-scale mass and essentially decouples
from particle physics.  The low-energy effective theory for the
SM particles and their superpartners is just the globally
supersymmetric theory, with certain soft SUSY-breaking
parameters, as discussed in the previous section.  If we choose
$M_{\rm SUSY} \sim 10^{11}$~GeV, the soft parameters are all
${\cal O}(M_{\rm Weak})$, as desired.

In general, the precise values of these parameters are sensitive
to the details of physics at the Planck scale.  If, however, one
makes the further assumption that the so-called K\"ahler potential
takes a certain canonical form (this is the reason for the
qualifier ``minimal" in minimal SUGRA), the soft parameters
reduce to the following simple set: there is one common SUSY-breaking
scalar mass parameter, $m_0$, and one bilinear and one trilinear
scalar coupling, $B_0$ and $A_0$.  It is also customary to assume
that an underlying grand unification yields a universal gaugino
mass, $\mhf$. Aside from the SM parameters, the model is then
completely specified by the parameters
($\mu$, $m_0$, $\mhf$, $A_0$, $B_0$).

We stress here that the minimal SUGRA (mSUGRA) model is not a
fundamental theory.  It is simply a low-energy effective theory,
valid below some very high scale, $M \sim M_{\rm GUT} - M_{\rm Planck}$.
Indeed, the simple structure of the soft SUSY-breaking parameters
holds only at this super-high scale.  As discussed in Sec.~I,
phenomenological analyses require that the masses and couplings
be evolved down to the weak scale.  This leads to a diverse (but
highly constrained) spectrum of sparticle masses which serves
as a test of the assumptions that have been made.

For ``practical supersymmetry,'' one can forget that the mSUGRA
model was derived using supergravity.  All that supergravity did
was to generate a set of soft SUSY-breaking parameters at the
ultra-high energy scale.  The additional ``minimality assumption''
fixed universal boundary conditions for these parameters.  This
is, in fact, {\it not} a general property of supergravity models,
and can only hold if the high-energy dynamics obeys an additional
approximate global symmetry, such as a $U(n)$ symmetry for the $n$
matter and Higgs supermultiplets.  This symmetry is broken by the
superpotential interactions, so the supergravity justification of
the universal boundary conditions must be regarded with care.

Regardless of its origin, the mSUGRA model provides an attractive
and economical framework for phenomenological studies.  A
particularly nice feature of this model is that it gives rise
to the radiative breaking of electroweak symmetry.  Even though
all scalars have a common mass at the high scale (which we take
to be $M_{\rm GUT} \simeq 2\times 10^{16}$~GeV), contributions from
the top-quark Yukawa interactions to the renormalization group
evolution drive $m_{H_u}^2$, the mass squared of the Higgs field
responsible for the masses of the up-type fermions, to negative
values.   This triggers the dynamical breakdown of electroweak
symmetry.

The desired symmetry breaking pattern is obtained over a wide (but
not complete) range of model parameters.  Assuming that this is
indeed the origin of electroweak symmetry breaking, one can use the
observed value of $M_Z$ to determine $\mu^2$, and eliminate the
parameter $B_0$ in favor of $\tan\beta = v_u/v_d$, the ratio of
up- to down-type vacuum expectation values (vevs), so that all
sparticle masses and couplings are fixed by the following
parameter set:
\begin{displaymath}
(m_0, \mhf, A_0, \tan\beta, \sgn\mu)
\end{displaymath}
plus the SM parameters.  It is remarkable that this framework
is consistent with all experimental data and also with the grand
unification of the gauge couplings, as measured by LEP experiments.
As a bonus, the LSP, which is frequently the lightest neutralino,
is a viable candidate for cosmological dark matter.

The mSUGRA model has several generic features which are
important for phenomenology.

\begin{itemize}

\item The first two generations of squarks and sleptons
are separately (almost) degenerate in mass.  This degeneracy
ensures consistency with the suppression of flavor changing
neutral currents.  Furthermore, the squarks are always heavier
than the corresponding sleptons.  Finally, the sleptons are
much lighter than the squarks only if $m_{\tq} \simeq m_{\tg}$.

\item Squarks (of the first two generations) are never much lighter
than gluinos:  if squarks and gluinos are heavier than about
200~GeV, $m_{\tq} \gsim 0.8m_{\tg}$.

\item Typically, $\mu$ is much larger than the electroweak
gaugino masses, so the lighter chargino and the two lightest
neutralinos tend to be gaugino-like, while their heavier
siblings are mostly Higgsino-like.

\item The lightest neutral Higgs boson is approximately the
Higgs boson of the SM.  It must be lighter than about 130~GeV.
The other Higgs bosons are typically much heavier.

\end{itemize}

\noindent
It should be kept in mind that there are some exceptions to
these rules, as discussed by the Theory Subgroup \cite{THEORY}.

It is important to remember that, despite its apparent success,
the mSUGRA framework rests on untested assumptions about the
symmetries of high-scale dynamics.  It is possible -- indeed
probable -- that these assumptions may ultimately prove to be
incorrect.  It is important, therefore, to study alternative
scenarios where one (or more) of the assumptions is relaxed.
It is equally important to examine how these assumptions can
be directly tested in future experiments once supersymmetry is
discovered.  Both these issues were addressed by our group
during the course of the Snowmass Workshop.

\subsection{Modifications of the mSUGRA Model}

The mSUGRA framework provides a very constrained pattern of
sparticle masses and couplings.  This pattern is, however,
sensitive to details about the physics at the high scale, so
a measurement of sparticle properties can yield information
about symmetries of physics at a scale far beyond what might
be accessible at accelerators.  We review here some of the
extensions that were examined at this Workshop, and refer
to Ref. \cite{THEORY} for details.

\subsubsection{New Contributions to Scalar Masses}

Naively, one would think that if additional gauge symmetries
(beyond the SM) are broken at a very high scale, the effects
of these interactions should be suppressed at low energy.  The
pattern of sparticle masses provides a simple counterexample
to this intuition.  The point is that if sfermions (or, for that
matter, the Higgs bosons) are not singlets of this new gauge
group, supersymmetry generates quartic interactions between pairs
of ordinary fields and pairs of fields whose vevs break this
symmetry.  When these fields are replaced by their vevs, these
quartic terms provide a new source of mass for observable-sector
fields.  Such interactions can alter the masses of the squarks,
sleptons and Higgs bosons (and indirectly, of other sparticles).

During the Workshop, the Theory Subgroup studied a test case
where the extra gauge symmetry was $U(1)_{B-L}$.  They showed
that the usual hierarchy between slepton masses can become
inverted; {\it i.e.} $m_{\tell_L,\tnu} < m_{\tell_R}$ is possible.
This would modify the leptonic branching fractions of the
charginos and neutralinos, and hence the multilepton signals
from cascade decays of gluinos and squarks.  For more details,
and for a discussion of the impact on the phenomenology of a
particular model, we refer the reader to Ref. \cite{THEORY},
and references therein.

\subsubsection{Non-Universal Unification-Scale Boundary Conditions}

We have already stressed that the universal boundary conditions
adopted in the mSUGRA picture result from untested assumptions about
the nature of high-scale physics.  This motivated the Theory Subgroup
to consider models with non-universal gaugino and scalar masses.
Non-universal gaugino masses may arise in models in which a non-minimal
gauge-field kinetic term is induced by the vev of a superfield that
is charged under the GUT group.  Alternatively, there are
superstring-motivated models where the boundary conditions are different
from those of the mSUGRA scenario.

In general, the pattern of sparticle masses is sensitive to model
parameters.  Here, we will confine ourselves to noting that the
phenomenology at the LHC and NLC can be very different from that
expected in mSUGRA.  An extreme example is provided by the
string-motivated O-II scenario in which the chargino is essentially
degenerate (to a fraction of a GeV for the particular parameter choice)
with the LSP and also to within 10 GeV of the gluino.  The ``visible''
decay products of the chargino may be too soft to be detected, at even
the NLC.  (However, this signal can be detected via the $e^+e^- \to
\tilde{\chi}_1^+\tilde{\chi}_1^-\gamma$ reaction, where only a high
energy photon is observed.)  Very soft jet
cuts would be required to dig out the gluino signal at the LHC
\cite{GUNIONLHC}.  While this scenario might be contrived, it exemplifies
the importance of general purpose detectors to cope with unexpected
surprises.

\subsubsection{Data Driven MSSM Scenarios}

Guided by a handful of unusual events in the Fermilab Tevatron
experiments, as well as a small deviation in LEP experiments, some
members of the Theory Subgroup attempted to explain these anomalies
in terms of supersymmetry.  They used selective subsets of the data
to guide them to model parameters.  For more information on this
data-driven approach, we refer the reader to the literature, where
the values of parameters ``as extracted  from the data'' are
documented \cite{BARNETT,KANE}.

\subsubsection{$R$-Parity Violation}

A disconcerting feature of supersymmetry is that it is possible to
introduce renormalizable interactions consistent with the SM gauge
symmetries which violate baryon ($B$) and/or lepton ($L$) numbers
(and also $R$-parity).  A theory with all such interactions would
be a phenomenological disaster since the proton would decay at the
weak rate.  In the absence of any deeper understanding of this issue,
we are forced to impose a discrete symmetry to stabilize the proton.
The canonical choice is the conservation of $R$-parity, but $B$ or
$L$ conservation do this job equally well.  The phenomenology is,
however, sensitive to this choice.  If $R$-parity is violated,
the LSP is unstable and can decay into SM particles; hence
the missing-energy signal is greatly altered.  The impact of this scenario
on the reach of the LHC was examined at this Workshop, and is
discussed in the next section.

\subsubsection{Gauge-Mediated Symmetry Breaking}

Recently, there has been a resurgence of interest in models where
SM gauge interactions, not gravity, act as the messengers of supersymmetry
breaking.  Supersymmetry is assumed to be dynamically broken in a hidden
sector of the theory, which is coupled to a ``messenger sector'' (with
a mass scale $M$) by a new set of gauge interactions.  Some particles in
the messenger sector are assumed to have SM gauge interactions.
Ordinary gauge interactions then induce soft SUSY-breaking masses for
the superpartners of the quarks and leptons, as well as for those
of the gauge and Higgs bosons.

An important difference from the mSUGRA model is that the
sparticles receive masses according to their gauge couplings.  In
the simplest models of this class, the gaugino masses satisfy the
``mass unification condition'' of Sec.~I, even though there need
be no grand unification.  Similarly, the
squarks are much heavier than sleptons because of their strong
interactions, while $m_{\tell_L}$ is substantially larger than
$m_{\tell_R}$ because the $SU(2)$ gauge coupling is larger than
the hypercharge gauge coupling.

Typically, gauge-mediated models are rather constrained, and for
the minimal model, sparticle masses and mixing patterns are
determined by the parameters ($\Lambda$, $M$,
$\tan\beta$, $\sgn\mu$).  Here $\Lambda = M^2_{\rm SUSY}/M$
determines the overall scale of the SUSY
spectrum, while the boundary conditions on the soft SUSY-breaking
parameters are specified at the scale $M$.  The soft parameters
are evolved from $M$ to the weak scale, just as in the mSUGRA
model.  The $A$ parameter at the scale $M$ is small because it is
induced at two loops.  As before, $\mu$ is fixed via the radiative
breaking of electroweak symmetry.

A very important feature of these models is that the SUSY-breaking
scale (which determines the gravitino mass) can be smaller than the
$10^{11}$~GeV of the SUGRA framework.  If this scale is $\sim$ 100
TeV, the gravitino mass is a fraction of an electron-volt!  As has
been pointed out by Fayet \cite{FAYET}, the longitudinal components
of such superlight gravitinos do not decouple, so the lightest
neutralino (which is typically the next-to-lightest supersymmetric
particle) can decay into a gravitino and a photon (or a $Z$ or Higgs
boson) via its photino (zino or Higgsino) component at a rate that
might be relevant for collider phenomenology.  This leads to novel
signatures for supersymmetry involving hard, isolated photons.  In
models with a larger messenger sector, $m_{\tell_R} < m_{\tz_1}$,
and the phenomenology can be quite different.  We refer the reader
to the report of the Theory Subgroup for further discussion of this
framework.  Several studies \cite{RECENT} have recently appeared
discussing the phenomenology of this class of models.

\section{THE SUPERSYMMETRY REACH OF FUTURE FACILITIES}

The SUSY reach of TeV33 \cite{BAER33,MRENNA,LOPEZ} and the LHC
\cite{BCPT,ATLAS} has been examined in several studies
prior to this Workshop.  Besides the standard missing-energy
signal, the cascade decays of gluinos and squarks also lead to
characteristic events with hard, isolated leptons plus jets and
missing transverse energy, $\eslt$.
In addition, the production and subsequent leptonic decays of charginos
and neutralinos gives rise to multilepton plus $\eslt$
events with low numbers of hadrons,
the most striking of which are trilepton events from $\tw_1\tz_2$
production.  This channel is especially important for SUSY discovery at
TeV33, where the production of gluinos and squarks (which are much heavier
than $\tw_1$ and $\tz_2$) is kinematically suppressed, and electroweak
chargino and neutralino production is the dominant source of SUSY
events.

\begin{figure}[b]
\leavevmode
\begin{center}
\resizebox{!}{6cm}{%
\includegraphics{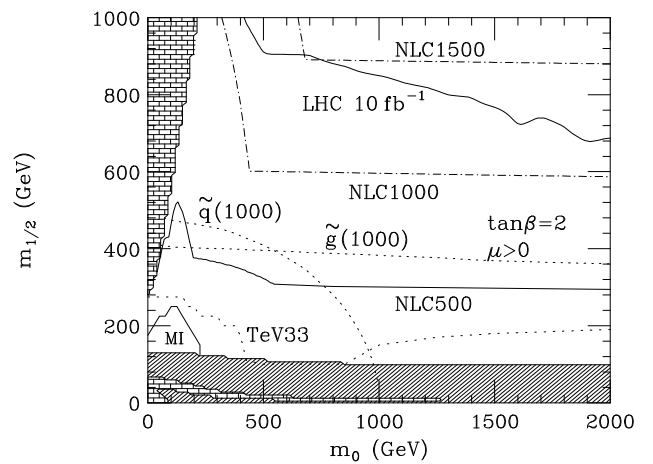}}
\end{center}
\caption{The SUSY reach of various facilities in the mSUGRA model,
for $\tan\beta=2$, $A_0=0$ and $\mu>0$.}
\label{fig1}
\end{figure}

At $e^+e^-$ colliders, it is well known that the discovery of charged
sparticles (and also the sneutrino, if it has visible decays) is relatively
straightforward for sparticle masses up to about half the beam energy.
The discovery reach is not contingent upon beam polarization, which as
we will see, is very useful for untangling the underlying production
processes.

Within the mSUGRA framework, the SUSY reach is largely determined by the
values of $m_0$ and $\mhf$, which fix the scales of the sparticle masses.
The $m_0-\mhf$ plane (for fixed values of $A_0$ and $\sgn\mu$) provides
a convenient panorama for comparing the reach of various facilities, as
shown in Fig.~1.  (The figure corresponds to $\tan\beta=2$, $A_0=0$
and $\mu>0$.  Qualitatively speaking, the reach is only weakly sensitive
to this choice.)

In the figure, the bricked (hatched) region is excluded by theoretical
(experimental) constraints \cite{BAER33}.  The region below the lines
labeled MI and TeV33 is where experiments at the Tevatron should be
able to discover SUSY, assuming an integrated luminosity of 2 and
25 fb$^{-1}$.  The discovery region is a
composite of several possible discovery channels, although the $\eslt$
and clean $3\ell$ channels dominate the reach.  To help orient
the reader, we have also shown contours for gluino and squark masses
of 1~TeV.

The upper solid line of Fig.~1 shows the boundary of the corresponding
region at the LHC \cite{BCPT}; it essentially coincides with the discovery
region in the $1\ell\ +$ jets $+\eslt$ channel.  Similarly, the solid line
labeled NLC500 denotes the reach of the NLC operating at $\sqrt{s}=0.5$
TeV, as obtained using ISAJET \cite{BMT}.  It consists of three parts:
the horizontal portion at $\mhf \sim$ 300 GeV essentially follows the
$m_{\tw_1}=250$~GeV contour, while the rising portion below $m_0=200$~GeV
follows the $m_{\te_R}=250$~GeV contour.  (The reach drops when $m_{\te_R}
\simeq m_{\tz_1}$ because the daughter electron becomes too soft.)
An observable signal from $e^+e^- \to \tz_1\tz_2$ makes up the intermediate
portion of the contour.  The dashed-dotted contours mark the boundaries
of the region where $\tw_1$ and/or $\te_R$ are kinematically
accessible at NLC1000 or 1500.
Although new backgrounds from two-to-three- and four-particle
production processes have not been evaluated, we believe that this region
closely approximates the reach of the NLC operating at these higher energies.

Several comments are worthy of note:

\begin{itemize}

\item Although TeV33 can probe an interesting region of parameter space,
there is a significant range of parameters, consistent with qualitative
upper bounds on sparticle masses from fine tuning arguments, where there
is no observable signal.  This can even be true if the chargino mass
is at its current experimental lower bound.  Nevertheless, it is important
to search directly for gluinos even in regions of parameter space
``excluded'' by the LEP experiments, since this exclusion requires
an assumption about gaugino mass unification that may prove to be
incorrect.  Note that the discovery reach of TeV33 is not overwhelmingly
larger than that of the Main Injector (MI).  Of course, if any signal
is discovered during the MI run, the larger data sample from TeV33 would be
extremely helpful in elucidating its origin.

\item The LHC has a very large reach for supersymmetry and can see gluinos
and squarks up to and beyond 1.7~TeV (2.2~TeV if gluinos and squarks are
roughly degenerate).  This provides a significant safety margin over the
upper limits expected from fine-tuning arguments.  Furthermore, one expects
a SUSY signal in several channels over much of the region with $\mhf
\lsim 500-600$~GeV.  For comparison, we quote the reach obtained by the
ATLAS collaboration \cite{ATLAS}, using a very different analysis within
the MSSM framework: they obtained a reach of
$m_{\tg} = 1600\ (1050)$ GeV, $2300\ (1800)$GeV, $3600\ (2600)$ GeV
for $m_{\tq}= 2m_{\tg}$, $m_{\tq}=m_{\tg}$ and $m_{\tq}=m_{\tg}/2$,
respectively, assuming an integrated luminosity of 100 fb$^{-1}$
(1 fb$^{-1}$).

\item For the purposes of assessing the SUSY reach (and for this purpose
alone), we see that an $e^+e^-$ collider operating between $1-1.5$~TeV
has a reach similar to that of the LHC.
\end{itemize}

Our considerations have thus far been confined to the mSUGRA framework
where the undetected LSP's provide the benchmark $\eslt$ signal.  We
have, however, seen in Sec.~II that if $R$-parity is violated, the
$\eslt$ signal may be greatly degraded.  At hadron colliders it is
reasonable to suppose that the worst-case scenario is when the LSP's decay
hadronically via operators which violate baryon number.  In this case,
not only is there no $\eslt$ from $\tz_1$, but the additional jets
from its decay degrade the lepton isolation, and so reduce the cross
section for the multilepton signals.  Indeed, it has been shown \cite{TEVRPV}
that gluinos and squarks as light as 200~GeV might escape detection
at the MI.  This led the LHC Subgroup \cite{LHC} to examine how the
SUSY reach is affected in such a scenario.  Assuming that the
$R$-parity violating interactions do not affect SUSY production
or the decays of any sparticle other than the LSP, it was shown
\cite{LHCRPV} that with a data sample of 10 fb$^{-1}$, experiments
at the LHC can observe gluinos and squarks with masses beyond 1~TeV,
even with cuts designed to detect SUSY as realized in the mSUGRA
model.

Finally, we mention that while the reach of
the NLC was not examined within the context of this scenario, we
expect that in the clean $e^+e^-$ environment, these signals would
give rise to very spherical events that would likely be
detected.  In fact, the SUSY reach may even be larger than in the
mSUGRA framework because the pair production of LSP's  would lead
to visible signals (assuming a sufficient cross section).

\section{DETERMINATION OF THE ORIGIN OF NEW PHYSICS SIGNALS}

An important goal of this study was to determine methods for
establishing whether an observed signal is of supersymmetric origin,
and further, to examine whether the new data are compatible with
mSUGRA or some other supersymmetric framework.
One would also like to know whether it is possible to determine
the underlying model parameters from the data.

Since it is not practical to do detailed signal studies over the
entire parameter space, even for the mSUGRA model (let alone its
variations), our strategy was to choose a few representative scenarios
to be studied by the TeV33, LHC and the NLC Subgroups.  To highlight
the different capabilities of the facilities, we made different
choices for the three subgroups.  However, to facilitate direct
comparison, we chose one common scenario in which the sparticle
spectrum was light enough to enable detailed study at each machine.
The mSUGRA parameters of this scenario are
\begin{displaymath}
(m_0,\mhf,A_0,\tan\beta,\sgn\mu)=(200,100,0,2,-1),
\end{displaymath}
where all mass parameters are in GeV.  For this choice, $m_{\tg} =
298$~GeV, with the first two generations of squarks about 20~GeV
heavier.  The slepton masses range between 206 and 216~GeV.  The
lighter chargino and the $\tz_2$ masses are 96 and 97~GeV, while
the LSP mass is 45~GeV.  The heavier charginos and neutralino masses
are between 260 and 270~GeV.  The lighter stop (sbottom) mass is
264 (278) GeV, while the corresponding heavier states are more
massive than the gluino.

For the parameters and sparticle masses in the other scenarios
studied at the Workshop, we refer the reader to the subgroup
reports \cite{LHC,TEV33,NLC} elsewhere in these proceedings.
All but one of these scenarios are within the mSUGRA framework.
We are aware that the various assumptions that underlie this
framework may ultimately prove to be incorrect.  Our attitude
is that this framework leads
to a rich variety of signals, and that even if it proves to be
erroneous, it serves as an excellent theoretical laboratory for
a broad comparison of the capabilities of the experimental
facilities for the discovery and analysis of
supersymmetry (and for that matter, other forms of new physics).

Not surprisingly, the strategy adopted by each of these subgroups was
very different.  We will, therefore, review the highlights of the results
of each subgroup separately, and conclude in the next section with a
comparison.

\subsection{Capabilities of the TeV33 Upgrade}

\subsubsection{Charginos and Neutralinos}

The TeV33 Subgroup studied trilepton signals, requiring that there
be no more than one jet in any event, for the common scenario as
well as for another mSUGRA point, with $m_0=100$~GeV and $\mhf=
150$~GeV (and with the other parameters as for the common point).
They estimate about 500 (425) signal events for the common (other)
point, assuming an integrated luminosity of 30 fb$^{-1}$.   They
find that for the case where $\tz_2$ decays via three-body modes,
it should be possible to measure $m_{\tw_1}-m_{\tz_1}$ by determining
the end point of the mass distribution of the closest pair of
oppositely-charged same-flavor dileptons in the event.  Although
they give no quantitative estimate of the precision, it seems
clear from the sharp edge in the $m(\ell^+\ell^-)$ distribution
shown in Fig.~2 of Ref. \cite{TEV33} that this mass difference
can be well determined.  Since $\tw_1$ and $\tz_{1,2}$ are
gaugino-like over a wide range of mSUGRA parameters, this measurement
directly yields $\mhf$.

The TeV33 study of the dilepton distribution for the other scenario
reveals the care that must be exercised in arriving
at conclusions from the data.  In this scenario, the sleptons are
lighter than the parent charginos and neutralinos.  Therefore the
charginos and neutralinos decay to real sleptons whose subsequent
decays lead to a trilepton final state.  The subgroup finds
that although $m_{\tw_1}-m_{\tz_1}$ is about 70~GeV (as compared
to 51~GeV for the common point), the dilepton mass spectrum cuts
off at roughly the same point as in the previous study.  By plotting
several other distributions, the TeV33 Subgroup can
distinguish between the scenarios.  Even at
the MI, where just $\sim$ 20 signal events are expected, their
likelihood analysis (see Fig.~4, Ref. \cite{TEV33})
shows that the chance of confusing these scenarios is just 9\%.
At TeV33, where there are many more events, the discrimination
power should be significantly increased.

\subsubsection{Gluinos}

The TeV33 Subgroup examined signals from gluino production for the common
scenario.  In this case, $B(\tg \to b\tb_1)=0.89$, and $B(\tb_1 \to
b\tz_2)=0.86$.  Since the leptonic decays $\tz_2 \to \ell^+ \ell^-
\tz_1$, with $\ell = e,\ \mu$, have a combined branching fraction
$\sim$ 0.33, the main
signal for gluino pair production consists of events with several
$b$-jets together with several leptons and $\eslt$.

In their study, the TeV33 Subgroup attempted to separate the SUSY
signal from SM and instrumental backgrounds by requiring at least
two isolated opposite-sign leptons of the same flavor, accompanied
by at least two tagged $b$-jets and $\eslt> 20$~GeV.  They assume
a $b$-tagging efficiency of 66\% and show that the dilepton pairs
in these events have a mass distribution which cuts off at more or
less the same position as that from the trilepton sample discussed
previously.  This serves as an independent confirmation of their
origin in $\tz_2$ decays.

The TeV33 Subgroup repeated the technique first
used by the LHC Subgroup at this Workshop to identify the gluino
origin of this signal.  The idea \cite{LHC,YAO} is that for events
where the dilepton mass is close to the end point of the
$m(\ell^+\ell^-)$ distribution, the dilepton system as well as
$\tz_1$ are at rest in the rest frame of $\tz_2$.  For a fixed
value of $m_{\tz_1}$, which they chose to be $m_{\tz_2}/2$,
the 3-momentum of $\tz_2$ can readily be constructed from the
observed momenta of the leptons.  The $\tz_2$ can be combined with
a tagged $b$-jet (at the cost of a combinatorial background) to
reconstruct $\tb_1$.  Finally, the reconstructed $\tb_1$ can be
combined with another tagged $b$-jet to yield the parent gluino.
If multiple $b$ plus multilepton events have their origin in
the gluino decay cascade, a scatter plot in the $m_{\tb_1}-\Delta
m$ plane (where  $\Delta m =m_{\tg}-m_{\tb_1}$ and
the masses are reconstructed
as described) would show a clustering at the proper values of
$\Delta m$ and $m_{\tb_1}$, with the rest of the plot being
populated by the incorrect assignment of jet and $\tz_2$ momenta.
{}From a fit to the projection which yields the $\Delta m$ distribution,
the TeV33 Subgroup claims $\Delta m=23.3\pm 1.2$~GeV, which should be
compared to the input value of 20~GeV.  The systematics associated
with this measurement, as well as with the corresponding
``measurement'' of $m_{\tb_1}$, have yet to be analyzed.

Although the subgroup was not able to complete an analysis of
the multijet signal in the $\eslt$ channel, they suggest that such
an analysis will be possible because preliminary estimates for the
common scenario indicate that there should be a 10$\sigma$ excess
above the background, even at the MI.  They anticipate (although
this study has not been performed) that it should be possible to
extract gluino/squark masses as well as the gluino/squark admixture
in the SUSY sample at TeV33.

\subsubsection{A Light Scalar Top}

The left- and right-handed scalar top quarks can have significant
mixing, so the lightest mass eigenstate ($\tt_1$) may be much lighter
than the other squarks.  The TeV33 Subgroup examined the possibility of
discovering $\tt_1$ at the Tevatron in a mSUGRA model with $(m_0,
\ \mhf,\ A_0,\ \tan\beta,\ \sgn\mu) = (200,\ 130,\ -400,\ 2,\ +1)$.  In
this model, the lighter of the two stop states has a mass of 140~GeV
and decays via $\tt_1 \to b\tw_1$.  Since the chargino decays via
$\tz_1 f\bar{f'}$ into SM quarks and leptons, the final states from
$t$-squark pair production have the same topology as those from
$t\bar{t}$ production; an important difference is that stop events
tend to be softer because a significant portion of the energy in the
event resides in the massive LSP's.

The TeV33 Subgroup examined prospects for detecting the scalar top in
the dielectron plus jets channel.  To separate the SUSY signal
from the $t\bar{t}$ background, they require that $\Sigma H_T=
\Sigma |p_T^{jet\ i}|+ |p_T^{elec\ 1}| < 250$~GeV and $B = |p_T^{elec\ 1}|
+ |p_T^{elec\ 2}| +$ $|\eslt| < 100$~GeV \cite{SENDER}, in addition
to the usual acceptance cuts.  Their analysis, which does not include
a complete detector simulation, predicts a signal of 142 events versus
a background of 118 for a data sample of 37 fb$^{-1}$.  The
subgroup anticipates that a complete detector simulation will lead to
an overall reduction in detection efficiency of both signal and
background by a factor of two, and a further reduction by 60\% if
both $b$-jets need to be tagged to reduce Drell-Yan and
misidentification backgrounds.  The subgroup suggests that an
integrated luminosity of 20 fb$^{-1}$ will be necessary for an
unambiguous detection of this signal \cite{TEV33}.

\subsection{Capabilities of the LHC}

In Sec.~III we saw that the discovery of SUSY is no problem at
the LHC, even in the unfavorable case where the LSP decays
hadronically into three jets.  We also noted that several
SUSY reactions can contribute to each final state, and
that there are detectable signals in several channels over
the range of parameters where sparticle masses are smaller
than 1~TeV.  Indeed, the dominant backgrounds for any
particular SUSY process are {\it other} SUSY processes.

An especially interesting development at this Workshop was the
demonstration that experiments at the LHC have the potential to
make precision measurements of some SUSY parameters.  Here,
we will focus on the comparison scenario for brevity, and only
allude to the results for other case studies performed at the
Workshop \cite{LHC}.  Further details may be found in Ref.
\cite{HINCH}.

\subsubsection{A Measure of Gluino or Squark Masses}

Gluinos or squarks are expected to be the most copiously produced
sparticles at the LHC.  Measurements of their masses are
complicated by the fact that every SUSY event contains at least two
undetected LSP's.  In previous work, a precision of 15-25\% has been
typically achieved \cite{OLDMASS} for sparticle masses below about
700~GeV.  A different technique was proposed at this Workshop, where
it was shown that distributions in the variable \cite{PAIGE}
\begin{displaymath}
M_{\rm eff} = |p_T,1|+|p_T,2|+|p_T,3|+|p_T,4|+\eslt,
\end{displaymath}
defined as the scalar sum of the transverse energies of the four
hardest jets plus $\eslt$, yield a measure of the mass scale for
gluinos/squarks, defined by $M_{\rm SUSY}=\min(m_{\tg},m_{\tu_R})$,
to a precision of about 10\%.  Here, the choice of $\tu_R$ to
represent the squark mass scale is arbitrary.  Since this method
appears to require only a moderately clean sample of SUSY events,
it should be possible to determine $M_{\rm SUSY}$ for gluinos/squarks
as heavy as 1.5~TeV in about three years of low luminosity LHC
operation \cite{PAIGE}.  The efficacy of the method was demonstrated
for squarks and gluinos of $900-1000$~GeV (cases B and C studied by
this subgroup).

\subsubsection{Precision Measurements}

Precision measurements of supersymmetric parameters are possible
at the LHC because of large data samples.  Indeed, for the
comparison scenario, one expects 13.5M SUSY events with just
10 fb$^{-1}$ of integrated luminosity.
This enormous data set permits very
stringent selection cuts to isolate relatively pure samples of
events from particular SUSY processes.

Consider, for example, gluino production.  As we saw in TeV33
subsection, the decay chain $\tg \to b\tb_1 \to bb\tz_2 \to bb
\ell^+\ell^-\tz_1$ ($\ell=e,\mu$) has a branching fraction of
$\sim$ 25\%, so that even with a tagging efficiency of 60\%
(and $c$ misidentification of 10\%) per $b$-jet, gluino pair
production leads to 272K events with

\begin{itemize}
\item 4 tagged $b$-jets, and
\item 2 pairs of opposite-sign like-flavor isolated leptons, one
from each $\tz_2$ decay,
\end{itemize}

\noindent
per 10 fb$^{-1}$ at the LHC.  If only one of the neutralinos is
required to decay leptonically, gluino production results in
a sample of about 694K almost-as-spectacular events, with

\begin{itemize}
\item 4 tagged $b$-jets,
\item 2 isolated opposite-sign, same-flavor leptons, and
\item 2 non-$b$ jets.
\end{itemize}

A clean gluino sample is obtained by selecting events with six or more
jets, at least three of which are tagged as $b$-jets, together
with a pair of opposite-sign isolated leptons of the same flavor.  The
measurement of the end point of the dilepton mass distribution yields
$m_{\tz_2}-m_{\tz_1}$.  This measurement is limited only by the absolute
calibration of the electromagnetic calorimeter, and is quoted to yield a
precision $\leq$ 50 MeV!  Within the mSUGRA framework, this leads to a
very precise determination of $\mhf$.  By comparing events with two or
four leptons and four tagged $b$-jets, it is claimed \cite{HINCH}
that the product $B(\tb_1\to b\tz_2)\times B(\tz_2 \to e^+e^-\tz_1)$
of branching fractions can be measured to be 14 $\pm$ 0.5\%.

In the cascade chain $\tg \to b\tb_1 \to bb\tz_2 \to bb\ell^+\ell^-
\tz_1$ ($\ell=e,\mu$), the momentum of $\tz_2$, and hence the mass
of $\tb_1$, can be reconstructed as described in the TeV33 subsection.
The LHC Subgroup showed that about 6K gluino and sbottom events
can be reconstructed in this way per year of low luminosity running
at the LHC.  They find that the mass difference $m_{\tg}-m_{\tb_1}$
can be measured to $\pm 2$~GeV and that the measurement
is rather insensitive to
the assumed value for $m_{\tz_1}$; the systematic error from jet
energy calibration and the assumed value of $m_{\tz_1}$ is not
quoted.

The authors of Ref. \cite{HINCH} attempt to reconstruct the masses
of the superpartners of the light squarks via their decays $\tq \to
q\tz_2$, which has a branching fraction of 10\%.  There is, of course,
an enormous background from gluino events which they try to reject
by vetoing events with tagged $b$-jets.  The mass reconstruction is
performed as for $\tb_1$ above.  Even with a 90\% vetoing efficiency,
their final event sample contains a significant fraction of $\tb_1$
events, which broadens their mass peak.  They estimate that the mean
squark mass can be measured to about 20~GeV.

The LHC Subgroup also attempted to identify sleptons
produced via the Drell-Yan process.  They select events with at least
three leptons (from the cascade decay of $\tell_L \to \ell \tz_2$).  They
apply a stringent jet veto (no jets with $E_T>30$~GeV), which is crucial
to suppress events from gluino and squark production, and to control
backgrounds from $t\bar{t}$ production, where a lepton from the decay
of a $b$-quark is accidentally isolated.  They find the resulting
event sample to be dominated by trileptons from electroweak gaugino
production, so it does not seem possible to isolate the slepton
signal.

Nevertheless, the rates for clean
trilepton events may be regarded as {\it predictions} of the mSUGRA
framework since, as argued below, it is possible to extract the relevant
parameters from the measurements of the hadronic data sample
discussed above.  While it should, in principle, be possible to extract
$m_{\tell_L}$ by reconstructing the $\tz_2$ in its decay and combining
it with an additional lepton, the technique does not work in practice
because the trilepton signal is dominated by events from $\tw_1\tz_2$
production.  Requiring at least four isolated leptons might eliminate
this problem, but then the event rate is too small to be feasible, at
least for the low luminosity LHC option.

\subsubsection{Extraction of mSUGRA Parameters}

At the common point, the measurements,

\begin{itemize}
\item $m_{\tz_2}-m_{\tz_1} = 52.36 \pm 0.05$~GeV ($1\sigma$),
\item $m_{\tg}-m_{\tb_1} = 20.3 \pm 2$~GeV ($1\sigma$),
\end{itemize}

\noindent
together with the measurement of the light Higgs boson mass,

\begin{itemize}
\item $m_h = 68.3 \pm 3$~GeV,
\end{itemize}

\noindent
that should be possible at LEP 2 (the $\pm 3$~GeV is a theoretical
error which reflects the uncertainty in the computation of $m_h$ in
terms of underlying parameters) imply that

\begin{itemize}
\item $\mhf = 99.9 \pm 0.7$~GeV,
\item $m_0 =200^{+13}_{-8}$~GeV,
\item $\tan\beta =1.95 \pm 0.05$,
\item $\sgn\mu =-1$.
\end{itemize}

\noindent
The parameter $A_0$, which mainly affects phenomenology through
the masses and mixing angles of third-generation sfermions,
is not determined but is constrained to
be larger than $-400$~GeV.  Since the trilepton rates from $\tw_1\tz_2$
and slepton production are insensitive to $A_0$, we now understand why
the cross section for clean trilepton events serves as a non-trivial
test of the model.  Another independent test is provided by a
measurement of the branching ratio product for the $\tb_1$ and $\tz_2$
decays discussed above.  (For this small value of $\tan\beta$,
$b$-squark mixing effects, which depend on the $A$-parameter, are
small.)

The LHC Subgroup argues that it should be possible to test consistency
with the gaugino mass unification condition as follows.  One first
extracts the unification scale from the measured values of the gauge
couplings.  Next, assuming that $\tz_{1,2}$ are predominantly
gaugino-like, one uses the measured value of $m_{\tz_2}-m_{\tz_1}$
and the gaugino unification condition to extract $\mhf$, which one
then evolves back to the weak scale to obtain $m_{\tg}$ and the
neutralino masses.  Even allowing
for a 5\% error in this calculation, it is non-trivial to find a
value of $m_{\tz_1}$ that simultaneously yields agreement with the
experimental values of $m_{\tz_2}$ and $m_{\tg}$.  Such a match can
be taken to be a test of gaugino mass unification.

\subsubsection{Other Case Studies}

Preliminary work for other cases of mSUGRA parameters was done at
the Workshop.  A more complete analysis appears in Ref. \cite{HINCH}.
In some cases, the LHC Subgroup showed that it is possible to reconstruct
a light Higgs, $h$, produced via cascade decays of gluinos and squarks,
through its $b\bar{b}$ decay.  The LHC Subgroup estimates that it should
be possible to measure the Higgs mass to about $\pm 1$~GeV.

Another case study (case A) involves sleptons lighter than $\tw_1$
and $\tz_2$, so two-body decays of these sparticles to sleptons
are allowed.  (The $\tz_2$ slepton decay competes with its decay to
Higgs bosons.)  In this scenario, squarks and gluinos are rather heavy,
so hard cuts on $M_{\rm eff}$ and the hardest jet can be made to
enhance the signal over the SM backgrounds \cite{PAIGE}.  Because the
$\tz_2$ now decays through a cascade of two-body decays, the kinematic
edge of the dilepton mass distribution occurs below
$m_{\tz_2}-m_{\tz_1}$.  We refer the reader to Ref. \cite{HINCH} for
details about the extraction of the slepton mass.

Ref. \cite{HINCH} also includes a study of the mSUGRA parameters for
other LHC scenarios.  Typically, $\mhf$ can be measured
to a precision of a few percent.  Even for the extreme case where
gluinos and squarks are as heavy as 1~TeV, $\mhf$ can be obtained to
within 10\%.  The precision with which other parameters can be
extracted depends on the scenario, but it is found that for the
most part, it is possible to obtain allowed ranges of $\tan\beta$
and $m_0$.  Sometimes two possible solutions are found; more
detailed measurements would be required to discriminate between
them.  In brief, these studies underscore the capabilities of LHC
experiments and contain the first steps towards an
effective measurement strategy at the LHC \cite{BCPT,HINCH}.

\subsection{Capabilities of the NLC}

It is well accepted that the discovery of new physics is relatively
easy in the clean environment of an $e^+e^-$ collider, provided
the machine is operated above threshold for this new physics and that
it has sufficient luminosity.  Furthermore, at these machines,
aside from a few exceptions such as the neutralinos of supersymmetry,
all sparticles with non-vanishing electroweak quantum numbers are
produced with comparable cross sections.

Electron-positron colliders, therefore, provide the potential for
discovery of a wide variety of new physics.  We have already
seen that the discovery of charged sparticles (and sneutrinos, if
they decay visibly) is relatively straightforward.  Neutralinos,
however, can escape detection.  Gauge invariance requires that they
couple to the $Z$ boson only via their Higgsino components,
which implies that their
cross section can be strongly suppressed if they are gaugino-like
and selectrons are heavy.  The range of mSUGRA parameters where
it should be possible to discover SUSY at the NLC via neutralino
signals is shown by the intermediate portion of the discovery
contour in Fig.~1.

If sparticles are indeed produced at the NLC, it is interesting
to ask what can be learned about their properties.  The precise
knowledge of the collision energy greatly facilitates the
reconstruction of events with undetected particles in the final
state.  The JLC \cite{JLC} group has pioneered a program where
they have shown that it is possible
to measure chargino, LSP and slepton masses to a precision of $\sim$
2\%, and further, to make other precision measurements which will allow
incisive tests of the assumptions that underlie the mSUGRA framework.
The JLC study ignores cascade decays since it is assumed that the machine
energy can be tuned so that only one new sparticle is produced.  This
study makes innovative use of electron beam polarization,
which was taken to be 95\%.

At this Workshop, the capability of the NLC was examined \cite{NLC}
including the effects of cascade decays as given by ISAJET.  The beam
polarization was conservatively taken to be 80\%.  The discovery
potential and precision measurements that might be possible at the
NLC with cascade decays have also been examined in Ref. \cite{BMT}.

\subsubsection{Beam Polarization}

Although not critical for SUSY discovery, the availability of beam
polarization plays a crucial role in sorting out SUSY signals at
the NLC.  First, the use of a right-handed electron beam greatly
reduces \cite{JLC} the main SM background from $WW$ production.
Second, it predominantly selects right-handed selectron processes,
as can be seen from Fig.~9 of the subgroup report \cite{NLC}.  Of
course, a left-handed electron beam is better suited to study the
$\te_L$ or charginos of the mSUGRA framework.

The main point here is that beam polarization
makes it possible to single out subsets of sparticle reactions.
Another important benefit of beam polarization stems from the fact
that it increases the  number of observables: this has been put to
good use for the extraction of model parameters \cite{JLC}, and as
we will discuss below, also for direct tests of SUSY \cite{FENG,NOJIRI}.

\subsubsection{Mass Measurements}

The energy spectrum of particles produced by the two-body decay of
spin-zero particles ({\it e.g.} the spectrum of electrons produced
via $e^+e^- \to \tell_R^+\tell_R^- \to \ell^+\ell^-\tz_1\tz_1$)
is, aside from detector acceptance effects, flat; a measurement of
the endpoints of this distribution leads to a precise determination
of the selectron and LSP masses.  The NLC Subgroup used this technique
for the comparison scenario with right-polarized electrons.  They
obtain best-fit values of 45.1 $\pm$ 1.5~GeV and 208.2 $\pm$
0.7~GeV for $m_{\tz_1}$
and $m_{\te_R}$, which may be compared with the input values of
44.5~GeV and 206.6~GeV, respectively.

Since this kinematic procedure does not depend on whether the
daughters of the parent spin-zero particle are stable, it can
also be used in the presence of cascade decays if a sample of
events from a single SUSY reaction can be isolated.  With a
left-handed $e^-$ beam, the process $e^+e^- \to \tnu_e\bar{\tnu_e}
\to e^-{\tilde\chi}_1^+ e^+{\tilde\chi}_1^- \to e^+e^-\mu^{\pm}
\tz_1\tz_1jj$, where one of the charginos decays to a muon and
the other into jets, provides an opportunity for a simultaneous
measurement of $m_{\tnu_e}$ and $m_{\tw_1}$.  Assuming an integrated
luminosity of 20 fb$^{-1}$ at NLC500, the extracted masses,

\begin{itemize}
\item $m_{\tnu_e} = 207.5 \pm 2.5$~GeV,
\item $m_{\tw_1} = 97.0 \pm 1.2$~GeV,
\end{itemize}

\noindent
compare well with the input values of 206.6~GeV and 96.1~GeV,
respectively.

The same technique was used previously \cite{BMT} to extract the
mass of the lighter stop and chargino for parameters such that
$\tt_1$ decays via $\tt_1 \to b\tw_1$.  In this study, it was
shown that the $t$-squark mass can be measured to about 6\%
at 90\% CL (and better if the chargino mass is independently
determined).

The measurement of sparticle masses is also possible when sparticles
decay via three-body decays.  The NLC Subgroup analyzed a sample of $4j$
events from chargino pair production (NLC point 5), where both charginos
decay hadronically.  They find that the end point of the $E_{jj}$
distribution (even with the correct assignments of jets) is not as sharp
as for two-body decays.  However, by requiring $M_{jj}$ to lie in a
narrow range (around 30~GeV in this study), it is possible \cite{BMT}
to simulate a two-body decay.  The NLC Subgroup used this technique
to fit the energy spectrum and find

\begin{itemize}
\item $m_{\tw_1} = 107.5 \pm 6.5$~GeV,
\item $m_{\tz_1} = 55.0 \pm 3.5$~GeV,
\end{itemize}

\noindent
for an integrated luminosity of 50 fb$^{-1}$,
as compared to the input values of 109.8~GeV and 57.0~GeV, respectively.

It should be mentioned that ISAJET, which is used for these simulations,
does not include either initial-state radiation or beamsstrahlung.
These effects distort the energy spectra and potentially degrade the
precision with which the end points can be determined.  The NLC Subgroup
studied \cite{NLC} these effects using SPYTHIA.  They conclude that
the few percent change in the spectrum can be readily corrected so
the precision of the mass measurements is not adversely affected.
However, these radiation effects can reduce the production cross
sections, so the statistical error may increase somewhat.

\subsubsection{Model Independent Analysis}

The NLC Subgroup also devised a strategy for analyzing a sample
of SUSY events without assuming any particular SUSY framework.
Although their analysis was performed for the common scenario,
the procedure is really quite general.  They start by envisioning
a machine which would first run at about 350~GeV (which, incidentally,
allows a study of the top threshold).  If no sparticles are
discovered, they would then increase the energy.

For the case at hand, however, chargino signals would be readily
discovered during the initial 350~GeV run.  Since the masses are
most precisely measured just above threshold, the NLC Subgroup advocates
{\it lowering} the machine energy to 250~GeV and making measurements
with both left- and right-polarized (80\% polarization) electron
beams.  The reason for this is that the relative importance of
the sneutrino-mass-dependent $t$-channel amplitude is sensitive
to the beam polarization.
With 20 fb$^{-1}$ of integrated luminosity,
the cross sections with an 80\% left- (right-)
polarized beam can be measured to a precision of 1.5\% (2\%).
These measurements then allow one to {\it predict} that
$m_{\tnu_e} < 250$~GeV.  Furthermore, the chargino and LSP masses
can be measured to a precision of 1\% following the methods
described above.

The next step is to run the machine above the sneutrino threshold,
{\it i.e.} at $\sqrt{s}=500$~GeV.  Note that $SU(2)$ gauge symmetry
dictates that $m(\te_L)$ cannot be very different, so it is reasonable
to expect $\te_L\bar{\te_L}$ production as well.  With 20 fb$^{-1}$ of
integrated luminosity and an 80\% right-handed electron beam, the masses
of $\tnu_e$, $\te_R$, $\tmu_R$ and $\te_L$ can be measured to a
precision of 2\%, 1\%, 1.5\% and 7\%, respectively.

\subsubsection{Is it Supersymmetry?}

The observation of particles with the spins and internal quantum
numbers expected in SUSY would be an extremely strong indication
that nature is supersymmetric.  However, the most convincing proof
would come from testing the supersymmetric relations between the
couplings of the various particles.  Note that unlike the mass
relationships, the tree-level SUSY relationships between various
dimensionless couplings are preserved even if supersymmetry is
softly broken --- for instance, the electron-selectron-photino
coupling is fixed by the electromagnetic interaction.  Indeed,
the situation parallels that in spontaneously broken gauge theories,
where coupling constant relationships are preserved even though
the mass relations implied by gauge symmetry are badly violated.

In practice, a program to directly test the SUSY relationships
is complicated by the fact that sparticles are generally
model-dependent mixtures of states with definite gauge quantum
numbers.  Just how one would verify such relations
depends on the values of SUSY parameters.  The original proposal
\cite{FENG} focused on the pair production of charginos which,
in the gaugino (mixed) region, allowed verification of equality
of the gaugino-sneutrino-electron (Higgs-Higgsino-gaugino) coupling
and the $W$-boson-neutrino-electron (Higgs-Higgs-$W$-boson) gauge
coupling to a precision of about 30\%.

A more precise test \cite{NOJIRI} was suggested at this Workshop
where it was shown that an accurate measurement of the cross section
and angular distribution of electrons produced via $e^+e^- \to
\te^+_R \te^-_R$, could, with 100 fb$^{-1}$ of integrated luminosity,
lead to a 2\% measurement of the electron-selectron-bino coupling.
Such a precise measurement begins to be sensitive to radiative
corrections, which, in turn, are sensitive to the masses of sparticles
which may not be kinematically accessible.

All the tests described here were carried out using parton-level
calculations assuming 100\% beam polarization.  It has yet
to be studied how well they survive a more realistic simulation
as well as more conservative assumptions about beam polarization.

\subsubsection{Testing Gaugino Mass Unification}

A direct test \cite{JLC} of gaugino mass unification can be carried
out at NLC if both $\te_R$ and $\tw_1$ are accessible and if a
right-handed polarized electron beam is available.  The idea is that
a measurement of the two masses, $m_{\tw_1}$, $m_{\tz_1}$, and
the two cross sections, $\sigma_R(\te_R\te_R)$, $\sigma_R(\tilde{\chi}_1^+
\tilde{\chi}_1^-)$, can be used to determine the four parameters,
$\mu, \tan\beta$ and $M_1, M_2$,
the two soft SUSY-breaking gaugino masses that
enter the chargino and neutralino mass matrices.  The precision with
which each of these parameters can be determined depends on the model.
The gaugino masses are well determined if the chargino has substantial
gaugino components, so the gaugino mass relation can be tested
to a precision of few percent, assuming an integrated luminosity
of 20 fb$^{-1}$ (50 fb$^{-1}$) at $\sqrt{s} = 350$~(500)~GeV.

\subsubsection{Tests of the mSUGRA Model and Determination of
Underlying Parameters}

The direct measurement of sparticle masses allows several non-trivial
tests of the constrained mSUGRA framework.  The most obvious of these
is the unification of scalar masses which can be directly tested from
the flavor independence of the masses for each variety of slepton.
There is no {\it a priori} reason why $m_{\te_R}=m_{\tmu_R}=m_{\ttau_R}$
(and likewise for the left-handed sleptons).  If the unification of gaugino
masses is independently confirmed, the universality of scalar masses
can then be tested since $\Delta m_{LR}^2=m_{\te_L}^2-m_{\te_R}^2$
is largely determined by $\mhf$.  Since $\te_L$ and $\te_R$ belong
to different SM multiplets, their masses are unrelated, and are indeed
quite different in the various frameworks discussed by the Theory
Subgroup.  In fact, a measurement of $\Delta m_{LR}$ might well serve
to distinguish between some of these models.

How well do the various precision measurements determine the SUSY
parameters?  For the comparison scenario, the NLC Subgroup finds
\cite{NLC}

\begin{itemize}
\item $\delta m_0 = \pm 2.7$~GeV,
\item $\delta \mhf = ^{+2.5}_{-1.0}$~GeV,
\item $\delta \tan\beta = ^{+0.17}_{-0.31}$,
\item $\sgn\mu = -1$.
\end{itemize}

\noindent
At the comparison point these parameters determine the masses
of the first two generations of squarks to be just over 320~GeV.
Direct detection requires that the NLC run at $\sqrt{s}= 800$~GeV.
The NLC Subgroup claims that the squark masses should be measurable
to a precision of about 10\%.  Finding squarks at the expected
mass would provide striking support of the mSUGRA framework.

It would also be interesting to examine whether it might be
possible to isolate a sample of top squark events and extract
information about the remaining parameter,
$A_0$.\footnote{At the fixed point of the top-quark Yukawa
coupling, the low-energy parameters are determined \cite{CARENA}
by $m_0$, $\mhf$ and $\tan\beta$.  They are essentially independent
of $A_0$.  In this case it is neither possible nor relevant to
determine $A_0$.}  Other
measurements which should be possible at the NLC, {\it e.g.} of
branching fractions of charginos and neutralinos, would provide
additional tests of this framework.

\section{CONCLUDING REMARKS}

Although the mSUGRA scenario provides an economic, consistent and
calculable framework for phenomenological studies, one must keep
in mind that it is based on untested assumptions about the symmetries
of high-scale dynamics.  In Sec.~II we saw several examples in
which alternative assumptions about these symmetries change the
sparticle mass and mixing patterns, and therefore expectations
about SUSY signals at colliders.

Recently developed models in which SUSY breaking is communicated
to the observable sector via gauge interactions are especially
interesting because they provide a serious alternative to the
canonical scenario where gravity communicates this information.  In
these gauge-mediated models, SUSY signatures may be quite different
\cite{RECENT} from those in mSUGRA: depending on the details of the
model and parameters, they may include multilepton plus multijet
events with additional hard photons, $\tau$-leptons, or other particles,
possibly with displaced vertices or kinked tracks.
It should also be remembered
that $R$-parity might not be conserved, in which case the usual $\eslt$
signal might not be observable.

Luminosity upgrades of the Tevatron will probe the regions of SUSY
parameter space that are most favored by (subjective) fine-tuning arguments.
The reach of the Tevatron with 2 fb$^{-1}$ and 25 fb$^{-1}$ of integrated
luminosity is illustrated in Fig.~1.  It could well be that the first
SUSY signals will be found at the Tevatron, which should be operated
to accumulate as much data as can be handled by the detectors.  But
even with 25 fb$^{-1}$ of integrated luminosity, there is a substantial
region of parameter space where gluinos and squarks are lighter than
1~TeV, and there is no detectable signal.

Whether or not weak-scale SUSY exists will be decisively answered at
the LHC.  Within the mSUGRA framework, the reach of the LHC extends
considerably beyond what is generally regarded as ``acceptable'' if
SUSY is to stabilize a perturbatively coupled Higgs sector.  Even in
the experimentally difficult case where the LSP decays hadronically
within the detector, experiments at the LHC should be able to detect
the SUSY signal in several leptonic channels provided gluinos or squarks
are lighter than 1~TeV. A multitude of detectable signals is also
expected within the mSUGRA framework.

Even without the availability of beam polarization, the SUSY
reach of the NLC extends essentially to the beam energy, provided
there is sufficient integrated luminosity ($\sim$ 20 fb$^{-1}$ for
NLC500). An $e^+e^-$ collider operating at $\sqrt{s}=1300-1500$~GeV
would have about the same SUSY reach as the LHC.  NLC500 would be
a very interesting machine because it would be guaranteed to find
at least one of the Higgs bosons of SUSY (or the Higgs
boson of the SM) if Higgs-sector interactions are perturbative up
to the unification scale.  Indeed, if no Higgs boson is discovered
there, many accepted ideas would need to be rethought, since we
would be forced to conclude that perturbative unification
requires additional new physics at a rather low scale.

However, it is possible for
NLC500 to miss the direct detection of SUSY altogether.
Quite likely, the detailed exploration of the rich spectrum
of SUSY particles will require higher energy. In planning the NLC
design, it is {\it crucial} for the energy of the machine to be
upgradeable to $1-1.5$~TeV.  This would ensure that the facility
could completely explore the new physics at the TeV scale.

The Supersymmetry Working Group
examined other information which might be gleaned
from the data and which might serve to clarify the nature of the
new physics, discriminate between various theoretical models,
and lead to a determination of the underlying model parameters.
At the Tevatron, the clean trilepton signal from $\tw_1\tz_2$
production permits a measurement of the mass difference between
the two lightest neutralinos.  It also allows one to distinguish
between models where the neutralinos decay to real sleptons and
those where three-body decays dominate.  The TeV33 Subgroup argued
\cite{TEV33} that a measurement of $m_{\tg}-m_{\tb_1}$ might also
be possible.  The high luminosity of TeV33 is essential for these
measurements.

The LHC and NLC Subgroups developed various tests of the mSUGRA
framework and a set of methods for measuring the underlying parameters.
The best method depends on what the parameters turn out to be.  For
the comparison scenario, it was shown that the mSUGRA parameters
(other than $A_0$) can be extracted with comparable precision at
both facilities.  The LHC Subgroup obtained a better precision on
$\mhf$ and $\tan\beta$, while $m_0$ was more precisely measured at
the NLC.

At first sight it seems surprising that precision measurements are
possible at the LHC.  The point, however, is that the enormous size
of the SUSY data sample permits hard cuts which isolate relatively
pure samples of events in each SUSY channel.  These samples allow
the reconstruction of SUSY masses using kinematic constraints.
For the comparison scenario, this technique allowed for a measurement
of $m_{\tz_2}-m_{\tz_1}$ which was limited only by the systematic
error on the calibration of the electromagnetic calorimeter.
Squark/gluino mass measurements at the 10\% level should be
straightforward.  In some cases, the LHC Subgroup demonstrated that
mass differences between strongly interacting sparticles may also
be measurable.  It is worth noting that
a  10\% measurement of a squark-gluino
mass difference substantially constrains the model.

At the NLC, a few percent measurement of the chargino and slepton
(and perhaps also $\tnu$) masses will be possible if the particles
are kinematically accessible.  In view of current mass bounds from
the Tevatron, it would indeed be fortuitous if squarks are accessible
at NLC500.  Information about gluinos is generally very difficult
to obtain at an $e^+e^-$ collider.

The flavor-independence of slepton masses would furnish direct
evidence for the unification of scalar masses.  While the
feasibility of this measurement was demonstrated only at the NLC,
the unification of gaugino masses can be tested at both machines.
(The precision with which the equality between multi-electron
and multi-muon signal can be established at the LHC has not been
examined.  This would provide indirect evidence for slepton mass
universality.)  A direct test of the SUSY particle couplings appears
possible only at the NLC.

What if the mSUGRA turns out to be incorrect?  Since it is
possible to measure several sparticle masses at the
NLC, these data will point to the correct theoretical
picture.  At the LHC it should be straightforward to
{\it exclude} mSUGRA, or any other particular framework,
via a standard $\chi^2$ analysis.  However, the broader
issue of how to proceed from these data to a determination
of the correct theoretical picture is less obvious.

We have seen that experiments at the LHC and NLC do indeed
provide complementary information.  Together, they will not
only reveal supersymmetry if it exists at the weak scale,
but will also carry out detailed measurements of the
sparticle properties.   This information will provide
clues that may help us unravel the dynamics of supersymmetry
breaking, and further our understanding of how electroweak gauge
symmetry is broken.

We are grateful for the input from all the members of the
Working Group that made this report possible.
This research was supported in part by the U.S. DOE
contracts DE-FG03-94ER40833, DE-FG03-95ER40894 and U.S.
NSF grant NSF-PHY-9404057.

%
%

\end{document}